\documentstyle[epsf]{mn}

\title[]{{\sl XMM-Newton} observations of the eclipsing polar EP Dra}

\author[Ramsay et al]{
Gavin Ramsay$^{1}$, C. M. Bridge$^{1}$, Mark Cropper$^{1}$, K. O. 
Mason$^{1}$, F. A. C\'{o}rdova$^{2}$ \and W. Priedhorsky$^{3}$\\
$^{1}$Mullard Space Science Laboratory, University College London,
Holmbury St. Mary, Dorking, Surrey, RH5 6NT, UK\\
$^{2}$University of California, Riverside, CA 92521, USA\\
$^{3}$Los Alamos National Laboratory, MS D436, Los Alamos, NM 87545, USA}
\date{Received: }

\begin{document}
\outer\def\gtae {$\buildrel {\lower3pt\hbox{$>$}} \over 
{\lower2pt\hbox{$\sim$}} $}
\outer\def\ltae {$\buildrel {\lower3pt\hbox{$<$}} \over 
{\lower2pt\hbox{$\sim$}} $}
\newcommand{\ergscm} {ergs s$^{-1}$ cm$^{-2}$}
\newcommand{\ergss} {ergs s$^{-1}$}
\newcommand{\ergsd} {ergs s$^{-1}$ $d^{2}_{100}$}
\newcommand{\pcmsq} {cm$^{-2}$}
\newcommand{\ros} {\sl ROSAT}
\newcommand{\exo} {\sl EXOSAT}
\newcommand{\xmm} {\sl XMM-Newton}
\def\rchi{{${\chi}_{\nu}^{2}$}}
\def\uchi{{${\chi}^{2}$}}
\newcommand{\Msun} {$M_{\odot}$}
\newcommand{\Mwd} {$M_{wd}$}
\def\Mdot{\hbox{$\dot M$}}
\def\mdot{\hbox{$\dot m$}}

\maketitle

\begin{abstract}

We present {\sl XMM-Newton} observations of the eclipsing polar EP Dra
which cover nearly 3 binary orbital cycles. The X-ray and UV data show
evidence for a prominent dip before the eclipse which is due to the
accretion stream obscuring the accretion region. The dip ingress is
rapid in hard X-rays suggesting there is a highly collimated core of
absorption. We find that a different level of absorption column
density is required to match the observed count rates in different
energy bands. We propose that this is due to the fact that different
absorption components should be used to model the reprocessed X-rays,
the shocked X-ray component and the UV emission and explore the affect
that this has on the resulting fits to the spectrum. Further, there is
evidence that absorption starts to obscure the softer X-rays shortly
after the onset of the bright phase. This suggests that material is
threaded by an unusually wide range of magnetic field lines,
consistent with the suggestion of Bridge et al.  We find that the
period is slightly greater than that determined by Schwope \& Mengel.

\end{abstract}


\section{Introduction}

Polars or AM Her systems are accreting binary systems in which
material transfers from a dwarf secondary star onto a magnetic
($B\sim$10--200MG) white dwarf through Roche lobe overflow. They vary
in brightness over the binary orbital period, which is synchronised
with the spin period of the white dwarf. It has become clear that the
accretion flow in these systems is highly variable from one orbital
cycle to the next. In eclipsing polars, once the secondary star has
completely eclipsed the white dwarf, the accretion flow between the
stars is still visible for a short time. In the optical this eclipse
profile is seen to vary from cycle to cycle (eg HU Aqr, Bridge et al
2002 and EP Dra, Bridge et al 2003).

In the case of EP Dra, which has an orbital period of 104.8 min,
Bridge et al (2003) attributed the varying eclipse profile to a
variation in the amount and location of bright material in the
accretion stream. They also found a depression or `trough' like
feature starting at $\phi\sim$0.9 in some cycles. Bridge et al
proposed that the accretion material makes a wide accretion curtain
which feeds many magnetic field lines before accreting onto the white
dwarf. This should result in the X-ray pre-eclipse spectrum being
harder than the post-eclipse spectrum.

EP Dra was included in the {\xmm}-MSSL survey of polars (Ramsay \&
Cropper 2003).  The integrated bright phase fluxes in the soft and
hard X-ray bands have already been presented by Ramsay \& Cropper
(2004) who showed that it had a spectrum consistent with the standard
accretion model of Lamb \& Masters (1979) and King \& Lasota
(1979). Here we present energy resolved light curves and phase
resolved spectroscopy of EP Dra with an emphasis on determining how
the absorption varies through the orbital cycle.

\section{Observations}

EP Dra was observed using {\it XMM-Newton} (Jansen et al 2001) on 2002
October 18 ({\xmm} orbit 523), approximately two years after the
optical observations of Bridge et al (2003). A summary of the
observations is given in Table~\ref{tab:obs}. The EPIC-MOS (Turner et
al 2001) start time is earlier than that of the EPIC-pn (Str\"{u}der
et al 2001), and the observation is slightly longer in duration.
Observations with the Optical Monitor (Mason et al 2001) were carried
out in three filter bands: the $V$-band, UVW1 (2400$\--$3400\,\AA) and
UVW2 (1800$\--$2400\,\AA).

The source counts were extracted from an aperture of
29\,$^{\prime\prime}$ for the EPIC-pn and 44\,$^{\prime\prime}$ for
the EPIC-MOS, centred on the source, and the backgrounds were
extracted from annuli around the source aperture of radii
98\,$^{\prime\prime}$ for EPIC-pn and 147\,$^{\prime\prime}$ for
EPIC-MOS (the particle background increased toward the end of the
observation). The data were processed using the {\it XMM-Newton}
Science Analysis Software version 5.4. At its brightest point
($\phi\sim0.78$), the $V$-band magnitude is 16.6. This is comparable
with that of $V\sim17$ of Schwope \& Mengel (1997) and the optical
observations of Bridge et al (2003) indicating that EP~Dra was in a
similar, high, accretion state.

\begin{table}
\begin{center}
\begin{tabular}{llcr}
\hline
Instrument & Mode & Filter & Duration\\
\hline
EPIC MOS 1 & Partial window & Thin & 18005\,s\\
EPIC MOS 2 & Partial window & Thin & 18022\,s\\
EPIC PN	   & Large window   & Thin & 17670\,s\\
OM	   & Fast Mode & V	    & 4399\,s\\
OM	   & Fast Mode & UVW1   & 4400\,s\\
OM	   & Fast Mode & UVW2   & 4399\,s\\
OM	   & Fast Mode & UVW2   & 1800\,s\\
\hline
\end{tabular}
\caption{{\it XMM-Newton} observation summary for EP~Dra from 2002 October 18.}
\label{tab:obs}
\end{center}
\end{table}

\section{Light curves}
\label{sec:lightcurves}

\begin{figure*}
\begin{center}
\setlength{\unitlength}{1cm}
\begin{picture}(13,7)
\put(-2.,-3.8){\includegraphics{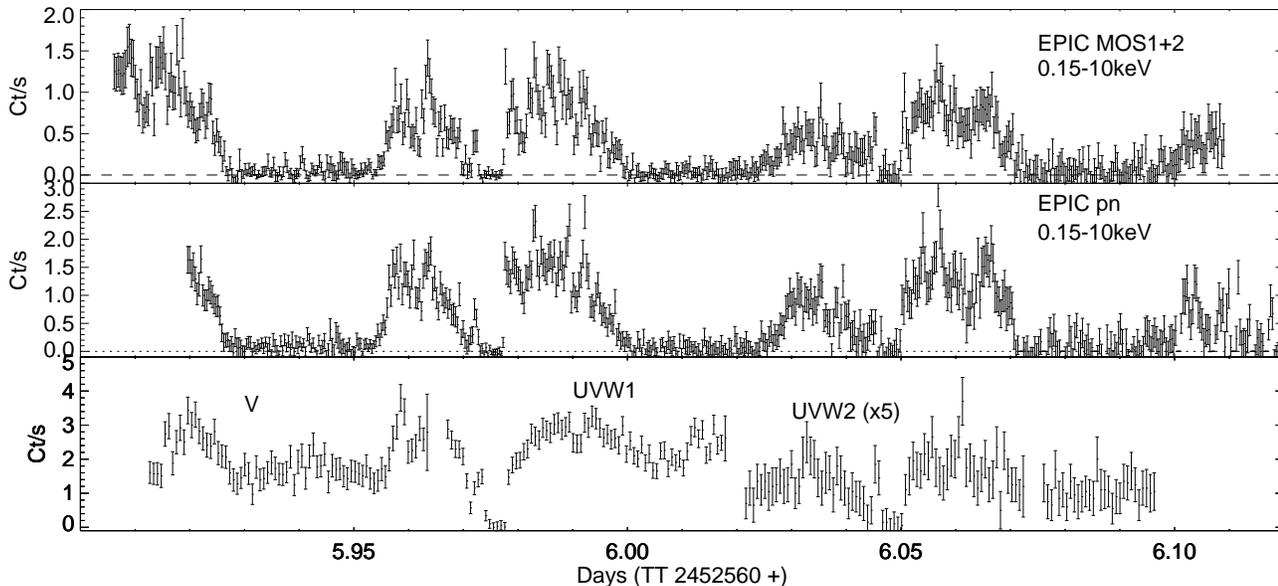}}
\end{picture}
\end{center}
\caption{The EPIC-pn, combined EPIC-MOS(1+2) and OM light curves for
EP~Dra. The X-ray light curves are in 30\,s time bins, in the energy range
$0.15\--10$\,keV. The $V$-band, UVW1/2 light curves are in 60\,s time
bins. The X-ray particle background is enhanced in the second half of
the observation.}
\label{lightcurve}
\end{figure*}

The X-ray light curves cover nearly three complete cycles of
EP~Dra. Figure~\ref{lightcurve} shows the light curve in the
0.15--10keV energy band for data taken using both the EPIC-pn and
combined EPIC-MOS(1+2), together with the OM data. The X-ray data is
of a much higher signal to noise and time resolution compared to the
{\sl ROSAT} observations of EP Dra presented by Schlegel \& Mukai
(1995) and Schlegel (1999).

\subsection{X-ray}

EP~Dra shows a distinct bright phase lasting $\Delta\phi\sim
0.6$. This indicates that the observable accretion region is located
in the upper hemisphere of the white dwarf, or if located in the lower
hemisphere it is spatially extended. The approximate centre of the
bright phase is cut by the eclipse of the accretion region and rest of
the white dwarf by the secondary star (the count rate at eclipse is
consistent with zero): the nature of the eclipse is discussed below.

The phase-folded light curves are split into two energy bands which
correspond to the energy ranges 0.15--0.5keV and 2--10keV (Figure
\ref{phase}). The hard X-rays emitted from the accretion region are
expected to be optically thin resulting in a light curve which is
quasi `top-hat' in form. The fact that the hard X-rays take $\sim$0.08
cycles to reach maximum indicate that the accretion region is extended
in azimuth on the white dwarf, or has significant vertical height. In
soft X-rays there is a steady decrease in flux shortly after reaching
maximum flux: by $\phi\sim$0.92 the flux is consistent with zero. In
hard X-rays there is no distinct decrease in flux until $\phi\sim$0.93
when there is a rapid drop in flux. This is the pre-eclipse dip seen
in a number of polars and is attributed to the accretion stream
obscuring our line of sight to the accretion region.

These results are reflected in the softness ratio (Figure \ref{phase})
which hardens as the bright phase progresses, until at the phase of
the main dip there is virtually no observed flux. This is consistent
with the proposal of Bridge et al (2003) that there is a wide curtain
of material feeding the magnetic field lines in EP Dra.

After the eclipse, where accreting material is not expected to obscure
the accretion region to the same extent as the pre-eclipse phase, the
energy-resolved light curves are similar to that expected. There are
some variations in the softness ratio, but this is probably due to
flickering behaviour which is most prominent at softer energies.

\subsection{UV and optical}

\begin{figure}
\begin{center}
\setlength{\unitlength}{1cm}
\begin{picture}(6,10)
\put(-1.5,-1.5){\includegraphics{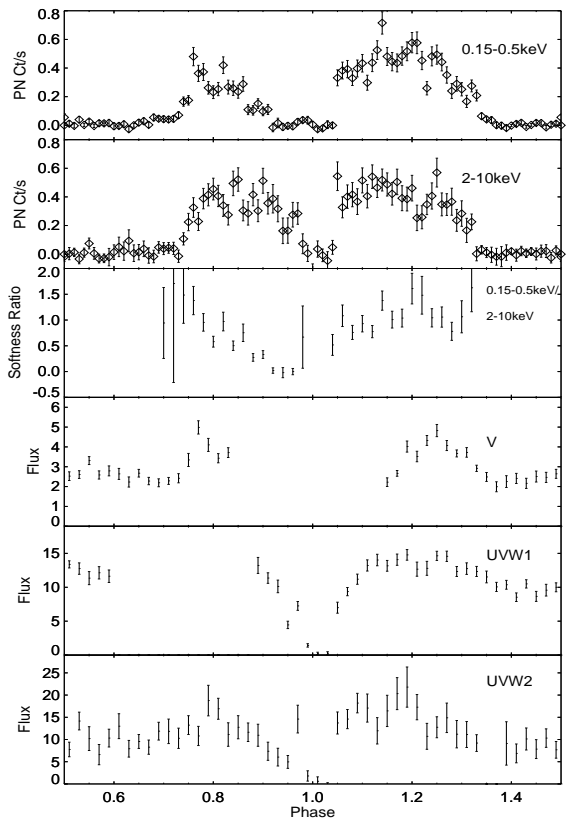}}
\end{picture}
\end{center}
\caption{The phase folded and combined EPIC and OM light curves. The
EPIC light curves are binned into 100 phase bins; $V$ band 40 sec
bins; UVW1 data 30 sec bins; UVW2 60 sec bins and the $B$ band 1 sec
bins. The $V$ band and UV fluxes are in units of erg s$^{-1}$
cm$^{-2}$ \AA$^{-1}$.}
\label{phase}
\end{figure}

The eclipse phase interval is seen in the UVW1 and UVW2 observations
but not in the $V$-band observation. During the eclipse, both UV
filters show a count rate consistent with zero during the
eclipse. Moreover, the pre-eclipse dip is reflected by a decrease in
the UV count rate in both UV filters (Figure \ref{phase}).

Outside of the eclipse the intensity variation is consistent with the
origin being a combination of the heated surface of the white dwarf
and the accretion stream. The $V$-band shows a strong rise in
brightness at $\phi\sim0.74$, which is consistent with the rise in
X-rays, and is most likely caused by cyclotron emission from the
post-shock stream material (Schwope \& Mengel 1997). The decrease in
brightness post-eclipse is more gradual, starting around $\phi\sim
0.3$ and ending around $\phi\sim 0.34$, again consistent with the end
of the bright phase in X-rays.

\subsection{The eclipse}
\label{the_eclipse}

\begin{figure}
\begin{center}
\setlength{\unitlength}{1cm}
\begin{picture}(6,5)
\put(-1.5,-0.5){\includegraphics{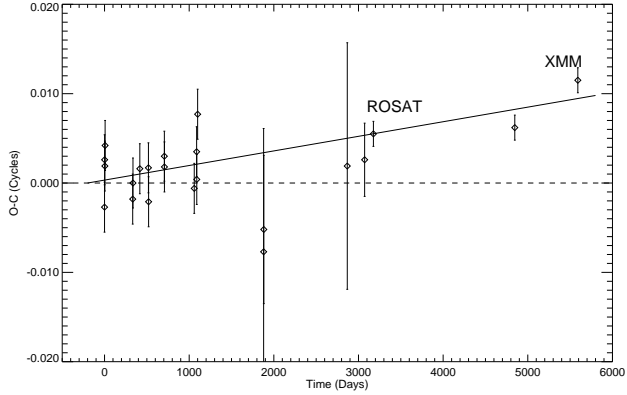}}
\end{picture}
\end{center}
\caption{The O-C diagram for the eclipse center based on data
presented in Remillard et al (1991), Schwope \& Mengel (1997), Bridge
et al (2003), the {\xmm} data here and {\ros} HRI data extracted from
the archive. We used the ephemeris of Schwope \& Mengel (1997) to
calculate the expected time of eclipse center. The fit to these times
is shown by a straight line. This suggests the period is slightly
greater than that estimated by Schwope \& Mengel.}
\label{o-c}
\end{figure}

The time system for {\xmm} data is Terrestial Time (TT). We therefore
convert the time zero of the ephemeris of Schwope \& Mengel (1997) to
units of TT.  When we fold the barycentrically corrected {\xmm} data
on the ephemeris of Schwope \& Mengel (1997), we find that the eclipse
is offset from $\phi$=0.0, with the eclipse centered at
$\phi\sim$0.01. This is consistent with the optical data of Bridge et
al (2003) which also shows an offset. To address this discrepancy, we
extracted {\ros} HRI data from the public archive and phase each event
from EP Dra on the ephemeris of Schwope \& Mengel (1997) (after
converting the {\ros} events times into units of TT). (We did not
include the {\ros} PSPC data since the source was at a large off-axis
angle and was faint).  We estimated the mid-point of the eclipse by
taking the mean point between the last event seen before eclipse and
the first event after eclipse. For the $B$ band data of Bridge et al
(2003) we estimated the mid-point from the original data rather than
their Table 1 which defines the eclipse mid-point as $\phi$=0.0 based
on the ephemeris of Schwope \& Mengel (1997). We also took the
mid-point of eclipse from Remillard et al (1991) and Schwope \& Mengel
(1997) after converting their times to TT time units and estimated the
difference between the observed time of eclipse and the expected time
using the ephemeris of Schwope \& Mengel (1997). We plot the residuals
in Figure \ref{o-c}: this suggests that the period was slightly
underestimated by Schwope \& Mengel (1997). We find a period of
0.072656272 days (an increase of 0.0011 sec) removes the residuals
shown in Figure \ref{o-c}.

\begin{figure}
\begin{center}
\setlength{\unitlength}{1cm}
\begin{picture}(6,11.5)
\put(-1.5,-0.5){\includegraphics{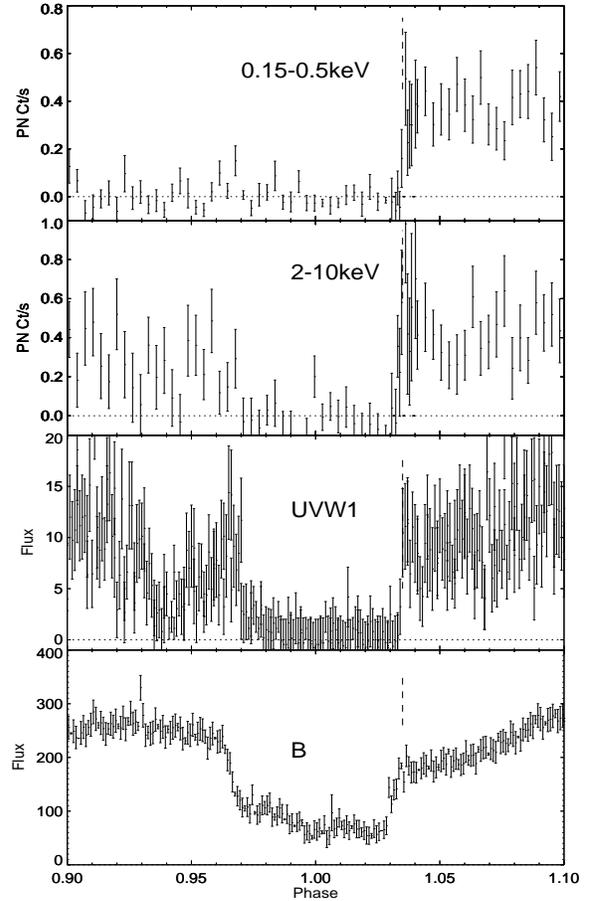}}
\end{picture}
\end{center}
\caption{The eclipse phase interval. From the top: the soft and hard
X-ray data from {\xmm}; the UVW1 data from {\xmm} and $B$ band data
taken from Bridge et al (2003). The top two panels have bins sizes of
20 sec, apart from the phase interval $\phi$=0.04--0.05 which is
binned into 5 sec bins. The UVW1 and $B$ band data 5 sec. The phase of
egress in the X-ray and UV bands ($\phi$=0.034) is shown as a dashed
vertical line.}
\label{eclipse}
\end{figure}

We show the {\xmm} and $B$ band data from Bridge et al (2003)
corrected so that the eclipse is centered on $\phi$=0.00 in Figure
\ref{eclipse}. In soft X-rays the phase of eclipse ingress is
complicated by the fact that before the eclipse the accretion stream
obscures the accretion region. In the $B$ band evidence for the
eclipse starts at $\phi$=0.961: this is in contrast to the UV and hard
X-rays where the eclipse is not seen for another $\sim$50 sec. In the
UV the eclipse ingress takes less than 5 sec, while in hard X-rays it
is more difficult to determine because of the lower count rate, but is
less than 20 sec. At eclipse egress, the $B$ band flux starts to
increase before the UV or X-ray emission. However, the $B$ band egress
finishes at the same phase at which the UV and X-ray emission
appears. In the UV and soft X-ray band the egress is very sharp,
taking $\sim$5 sec. In hard X-rays the rise time is more uncertain,
but similar. This demonstrates the difficulty of determining the
eclipse centers in different energy bands. In the optical the ingress
and egress marks the eclipse of the white dwarf, while in the X-ray
band it marks the eclipse of the accretion region which can be
anywhere on the surface of the white dwarf.

Using the standard Roche lobe geometry we can estimate the mass of the
white dwarf assuming a mass ratio, $q=M_{2}/M_{1}$. Using the standard
period-secondary-mass relationship for main sequence stars (cf Warner
1995) we estimate $M_{2}\sim$0.19\Msun for EP Dra. For inclinations of
78$^{\circ}$, 80$^{\circ}$ and 82$^{\circ}$ we find $M_{1}$=0.50, 0.68
and 0.95\Msun respectively. Schwope \& Mengel (1997) determine
$M_{1}$=0.43 \Msun from optical spectroscopy measurements.

\section{Spectra}
\label{spectra}

It is clear from Figure \ref{phase} that the pre-eclipse bright phase
is more absorbed than the post-eclipse bright phase. To place this on
a more quantitative level, we combined single and double events and
extracted spectra from the pre-eclipse ($\phi$=0.75--0.95) and
post-eclipse ($\phi$=0.05--0.25) bright phases. Only the first two
cycles were used as the particle background was much greater in the
third. The spectrum was fitted using the {\tt XSPEC} (Arnaud 1996)
spectral analysis package and the epn\_ff20\_sdY9\_thin.rsp response
matrix. For details of the X-ray spectra of polars observed in a high
accretion state in general and the energy balance of EP Dra in
particular see Ramsay \& Cropper (2004). Here we investigate the
specific question as to whether the absorption is greater before
eclipse compared to after it.

\begin{figure}
\begin{center}
\setlength{\unitlength}{1cm}
\begin{picture}(6,5.)
\put(-1.5,-0.8){\includegraphics{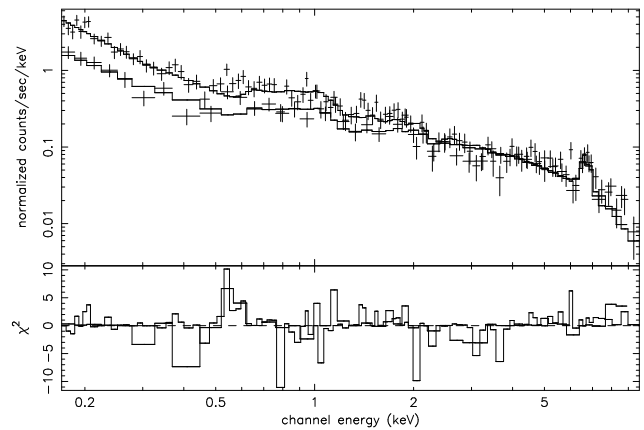}}
\end{picture}
\end{center}
\caption{The post-eclipse bright phase spectrum (upper set of crosses)
with best-fit model (upper solid line) and residuals; The pre-eclipse
bright phase spectrum (lower set of crosses) with best-fit model
(lower solid line) and residuals.}
\label{spec}
\end{figure}

We fitted the spectra using the stratified accretion emission model of
Cropper et al (1999) plus a blackbody component and a Gaussian at
6.4keV. We also added a neutral absorption component and a neutral
absorption component with partial covering. These parameters were
allowed to vary between the pre and post eclipse spectra, while the
emission parameters were tied. We fitted both spectra simultaneously.
A reasonably good fit was obtained (\rchi=1.48, 144dof). The residuals
are likely to be due to the fact that the spectrum is intrinsically
varying over the phases over which the spectra were accumulated (cf
the softness ratio, Figure \ref{phase}). We show the spectra in Figure
\ref{spec} and the absorption parameters in Table \ref{abspar}. This
shows that the pre-eclipse spectrum is more absorbed than the
post-eclipse spectrum.

\begin{table}
\begin{center}
\begin{tabular}{lrr}
\hline
 & Pre-eclipse & Post-eclipse \\
\hline
$N_{H,1}$ ($10^{20}$ \pcmsq) & 1.1$^{+0.5}_{-0.4}$ & 0.0$^{+0.2}$ \\ 
$N_{H,2}$ ($10^{22}$ \pcmsq) & 3.0$^{+0.6}_{-0.5}$ & 2.8$^{+0.6}_{-0.5}$\\
cvf$_{,2}$                        & 0.87$^{+0.01}_{-0.02}$ &
 0.79$^{+0.01}_{-0.02}$\\
\hline
\end{tabular}
\end{center}
\caption{The absorption parameters for X-ray spectra taken pre and
post eclipse. The first component is a neutral absorption component
while the second component is a neutral absorption component with a
partial covering fraction (cvf).}
\label{abspar}
\end{table}

\section{Absorption in the accretion stream}
\label{sec:absorption}

The soft X-ray light curve (Figure \ref{phase}) shows a general and
gradual decrease in flux from the start of the bright phase, leading
up to a prominent dip when the flux is consistent with zero. In hard
X-rays there is a sudden drop in flux to the pre-eclipse dip. This
implies that while the accretion flow becomes attached over a wide
range of magnetic field lines (Bridge et al 2003), there is a dense,
well defined, `core' of absorption. What makes EP Dra unusual is the
depth of the dip in the hard band, implying that the dense core has an
unusually high column depth.

Ramsay et al (2004) estimated the depth of the absorption column in GG
Leo and found that it was different depending on whether it was
determined using soft X-rays or in the UV. To estimate the absorption
column that would be needed to produce the observed drop in count rate
(the mean count rate in each band between $\phi$=0.93--0.97), we took
the pre-eclipse spectrum (\S \ref{spectra}) and increased the
absorption column to obtain the observed decrease in count rate for
soft X-rays, hard X-rays and the UV bands. For the X-ray data we
increased the neutral absorption model component, while for the UV
data we used the {\tt UVRED} model within {\tt XSPEC} (this is
necessary as the neutral absorber model component does not extend to
UV energies). We use the relationship between optical extinction and
total absorption column of Bohlin, Savage \& Drake (1978). We note
that this is likely to be a lower limit since the {\tt UVRED} model
assumes a large fraction of the UV absorption is caused by dust, which
is unlikely to be present in the accretion stream. We find we need a
column depth of 1--2$\times10^{21}$ \pcmsq in the UV bands,
8$\times10^{20}$ \pcmsq for 0.15--0.5keV and $1\times10^{23}$ \pcmsq
for 2--10keV bands: they are clearly inconsistent.

A solution may be that different absorption components should be
applied to the various emission components. This is not surprising
since the line of sight to the reprocessed emission site (which
produces a significant fraction of the soft X-ray flux, Ramsay \&
Cropper 2004) will be different to the line of sight to the post-shock
accretion region (which produces all of the harder X-rays). The UV
emission, will again have a different line of sight. However, almost
all investigators when fitting X-ray spectra of polars have modelled
their spectra assuming that the same absorption component can be used
to model the absorption of each of the emission components.

To investigate this further, we fitted the post-eclipse spectrum with
a model which applied separate absorption components to the blackbody
and the stratified emission region and also a model which applied the
same absorption model to both emission components (as in
\S\ref{spectra}). We applied one neutral absorption component to both
to account for interstellar absorption. We find that the former
(physically more realistic) model gives a better fit (\rchi=1.33,
92dof) compared to the latter model (\rchi=1.53, 94dof) at a
significance level of greater than 99.9 percent. Further, we find that
the blackbody emission component requires one absorption component,
while the stratified emission region component requires two absorption
components (with different column densities and covering fractions).

Ramsay et al (2004) determined the mass of the white dwarf in 3 polars
using {\xmm} data the same stratified emission model as used
here. They found that if they restricted the energy range to 3--10keV
rather than the full 0.2--10keV range they determined less massive
white dwarfs. This was due to the fact that absorption plays an
important role at lower energies and therefore there maybe some
degeneracy between the level of absorption and the slope of the
stratified emission model. However, Ramsay et al (2004) also noted
that the fits to the 3--10keV spectra were poorer compared to the fits
made using the full energy range and suggested that this may be due to
the absorption being more complex than applying the same absorption
model to both the blackbody and stratified emission models. Here we
find that separate absorption models should be applied to each
emission component. This would have implications for the resulting
mass of the white dwarf.

\section{Location of accretion stream material}
\label{accretionlocation}

As the accretion flow leaves the secondary star it falls under gravity
towards the white dwarf until the magnetic pressure of the white dwarf
balances the ballistic pressure of the in-falling material.  In
reality a treatment of the transition between the ballistic and the
magnetic dominated region is not simple. Different parts of the
accretion stream material will therefore satisfy this condition at
different points along the ballistic trajectory.  The actual location
of threading for a particular part of the accretion stream will depend
upon the density of the material, and the orientation of the magnetic
field lines relative to the in-falling material. This means that the
material will not all be be threaded by the magnetic field at the same
point (see, for example, Liebert \& Stockman 1985, Mukai 1988,
Heerlein, Horne \& Schwope 1999).

Evidence for material being coupled on to field lines covering a range
in magnetic azimuth has long been available (eg Ferrario, Tuohy \&
Wickramasinghe 1989). These {\xmm} observations show that in EP Dra,
the extent in phase of the soft X-ray absorption seen prior to the
eclipse centre indicates that the accretion stream attaches onto a
wide range of magnetic field lines of the white dwarf and is able to
obscure the X-ray emission site for a significant fraction of the
orbit. The stream does, however, have a dense and well defined
core. In contrast, in HU Aqr, Schwope et al (2001) show that material
is accreted over a narrower range of azimuth.

The fact that the main stream dip is located very shortly before the
eclipse, $\phi\sim$0.944 or an azimuth of $\sim20^{\circ}$, implies
that the most dense part of the stream becomes attached to the
magnetic field lines at a point relatively distant from the white
dwarf. We can compare this to the phase of the prominent dip seen in
HU Aqr. This has been extensively observed using X-ray instruments and
the azimuth of the main dip has been seen to vary from
34--50$^{\circ}$. This implies that the main accretion stream in EP
Dra is coupled by the magnetic field further from the white dwarf
compared to HU Aqr. In the standard accretion flow model the point at
which the stream gets attached by the magnetic field is the point at
which the magnetic pressure equals the ram pressure of the stream (cf
Mukai 1988).  The location of this point is thought to be relatively
weakly dependent on the mass transfer rate and the mass of the white
dwarf. It is more likely to be dependent on the magnetic moment
($\mu=BR_{wd}^{3}$) of the white dwarf and the angle that the magnetic
axis makes with the spin axis of the white dwarf, $\beta$.

The magnetic field strength of the white dwarf in HU Aqr is 36MG
(Glenn et al 1994) while in EP Dra it is 16 MG (Schwope \& Mengel
1997). For the coupling point to be further from the white dwarf in EP
Dra compared to HU Aqr this would imply that the white dwarf in EP Dra
is significantly less massive (larger radii and hence greater $\mu$)
compared to HU Aqr. In \S \ref{the_eclipse} we estimate
$M_{1}$=0.50\Msun for an inclination of $78^{\circ}$ -- this compares
with $M_{1}$=1\Msun for HU Aqr (Schwope et al 2001).

For high angles of $\beta$, the stream travels further before it feels
the effect of the magnetic field. Support for a high $\beta$ angle in
EP Dra may be found in the peak of the X-ray bright phase: with the
eclipse center fixed on $\phi$=0.0, the center of the hard X-ray
bright phase (defined as the mid point between the rise and end of the
bright phase) is $\phi$=0.02. This implies the accretion region is
behind the line of centers joining the two stars. This is unusual
since Cropper (1988) found that in most systems the bright phase leads
the secondary and has a mean azimuth of $\sim20^{\circ}$. High signal
to noise polarimetric data should be able to determine the co-latitude
of the accretion region.

\section{Conclusions}

We have presented X-ray energy resolved and optical/UV light curves of
the eclipsing polar EP Dra. They show a prominent dip in X-rays and in
the UV which is due to the accretion stream obscuring the accretion
region on the white dwarf. Our modelling of the X-ray spectra taken
before and after the eclipse of the secondary provides evidence that
accretion material attaches on to the magnetic field lines along a
wide arc along the ballistic trajectory supporting the suggestion of
Bridge et al (2003).

We find that the level of absorption needed to account for the
pre-eclipse absorption dip is much greater in hard X-rays compared to
softer X-rays.  We suggest that one factor which may cause this effect
is that the line of sight to the emission sites producing soft X-rays,
hard X-rays and UV is different. We test this by fitting such a model
to the X-ray data and find that a better fit is achieved compared to
using an absorption model which is applied to both emission
components. This has implications for fits to the continuum spectra of
polars.

\section{acknowledgments}

This is work based on observations obtained with {\sl XMM-Newton}, an
ESA science mission with instruments and contributions directly funded
by ESA Member States and the USA (NASA). We thank the referee for
helpful comments.

\end{document}